Success and Persistence in Science: The Influence of Classroom Climate

L. O. Dickie*[1], H. Dedic[2], S. Rosenfield[2], E. Rosenfield[3], R. A. Simon[4]

[1] John Abbott College, Ste-Anne-de-Bellevue, Quebec, Canada. [2] Vanier College and Concordia University, St-Laurent, Quebec, Canada. [3] Champlain College, St-Lambert, Quebec, Canada. [4] McGill University, Montreal, Quebec, Canada
*Corresponding author; **leslie.dickie@johnabbott.qc.ca**

Acknowledgement:
This work was supported by the Quebec Government through the program, Fonds Québécois de la recherche sur la société et la culture (FQRSC). The authors wish to thank the faculty and students who took the time to participate in this study and the many helpful discussions with the members of the research team.

Abstract

To better understand how student and faculty perceptions of the learning climate in science/mathematics classes influence success and persistence, we followed a cohort of 1425 academically able students who entered CEGEP in the fall of 2003. Students completed surveys in their first, second and fourth semesters. In the second semester 84 faculty members completed a similar survey. Faculty conceptions of teaching were identified using a framework developed by Scardamalia and Bereiter (1989). No significant gender differences in achievement were found. Self-efficacy declined over students' first semester as did affect towards science. Classes that students perceived as fostering their development had a positive impact on persistence and success while classes characterized as transmitting had a negative impact. Females were more likely than males to characterize a class as transmitting and to abandon science. Faculty members who had pedagogical training were more likely to create a fostering atmosphere in their classes.

Résumé

Afin de mieux comprendre comment la perception des élèves et des enseignants du climat pédagogique dans les classes de sciences/mathématiques influence le succès et la persévérance, nous avons suivi 1 425 étudiants possédant un rendement académique supérieur ayant commencé le CEGEP à l'automne 2003. Les étudiants ont complété un questionnaire durant leur premier, second et quatrième semestres. Au second semestre, 84 membres du personnel enseignant ont complété un questionnaire similaire. Les conceptions de l'enseignement des enseignants ont été identifiées à l'aide d'un cadre référentiel développé par Scardamalia et Bereiter (1989). Selon les résultats obtenus, aucune différence importante liée au sexe des candidats n'a été notée dans la performance académique de ces derniers. L'efficacité personnelle des élèves a baissé au cours du premier semestre, de même que leur apréciation des sciences. Le succès et la persévérance scolaires seraient affectés positivement lorsque les élèves perçoivent un soutien dans leur environnement d'études, contrairement aux cours caractérisés par la simple transmission du savoir qui, eux, ont un impact négatif. Les filles ont plus tendance que les garçons à décrire le climat pédagogique comme étant caractérisé par la transmission du savoir et, par conséquent, à abandonner les sciences. Les professeurs ayant reçu une formation en éducation avaient quant à eux davantage tendance à créer des environnements de soutien dans leurs classes.

Success and Persistence in Science: The Influence of Classroom Climate

Students, especially females, are abandoning science in university in disturbing numbers. Over the past twenty years the number of college bound students interested in science or engineering majors has dropped by 50% and moreover, as many as half of the students that do enter science programs transfer out (Pearson & Fechter, 1994). The current study found a steady attrition from high school to college to university (see Figure 1; Rosenfield et al., 2005).

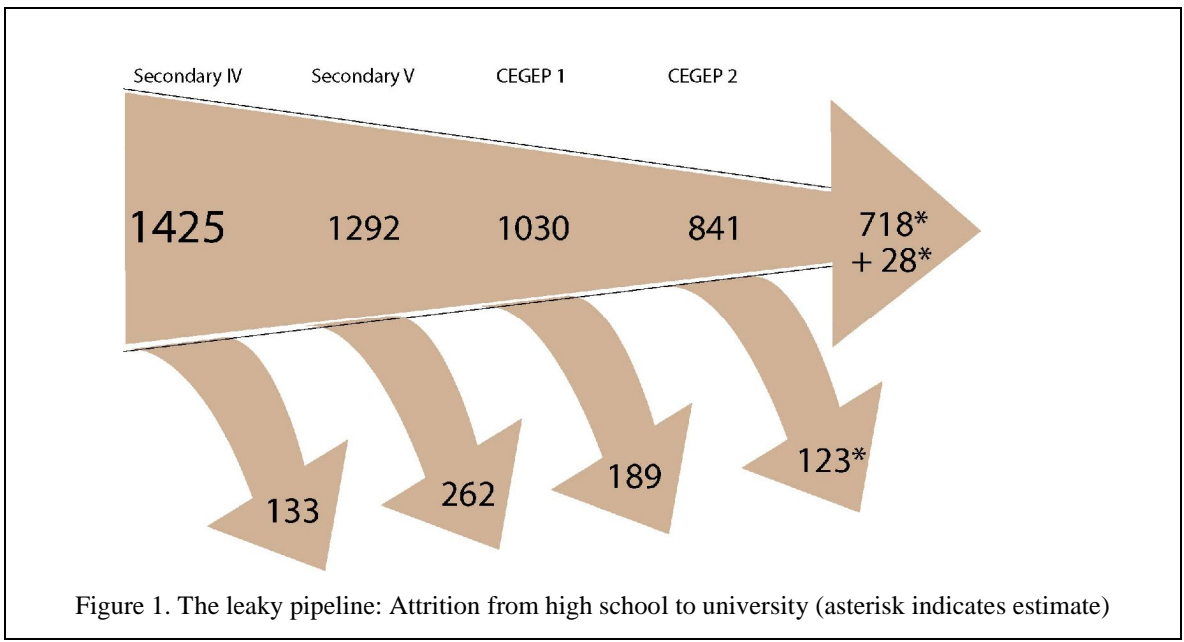

Figure 1. The leaky pipeline: Attrition from high school to university (asterisk indicates estimate)

This trend of fewer students graduating from the sciences is of concern because in a technological world, scientific literacy for all is essential if citizens are to make informed decisions on issues like global warming and the responsible use of the earth's resources (Grayson, 2006). Further, as we move away from the Industrial Age to the Knowledge and Communication Age, the wealth of each country increasingly depends upon its ability to educate an increasing percentage of its population as "knowledge-workers."

When students enter college classrooms their learning outcomes are influenced by contextual factors such as the course, the setting, their epistemological conceptions about learning and teaching, their prior knowledge of the subject and the perceived actions of the teacher. Students look to faculty for guidance about what to learn and how to learn: some of the direction is overt, some must be inferred from the actions and words of the teacher, and some from the reactions of the teacher to stimuli such as student questions. In turn the teacher's epistemological conceptions about learning and teaching and his/her beliefs about how students learn influence his/her actions (Haney, Lumpe, Czerniak & Egan, 2002). For example, whether the teacher views teaching as transmitting knowledge or as promoting conceptual change in students will affect the type and frequency of assessment tasks assigned, the degree of control the teacher maintains by either lecturing or enabling small-group work, and whether the teacher assumes responsibility for covering the material by providing handouts and library references or delegates part of this responsibility to the student by expecting them to be more active in finding their own resources (Kember & Kwan 2002; Vermut & Verloop, 1999).

Marginal use of inappropriate teaching methods by faculty is an important detriment to success and persistence in science (Seymour & Hewitt, 1997; Tobias, 1990): the problem does not start in college or university but has its roots in elementary school. Yager, Simmons and Pennik (1989) in a study of nine-year-old students, found that 75% of them felt positive about studying science, but that this percentage declined to 50% in 17-year-olds. Similarly, Davis and Steiger (1996) reported that student interest in studying science, mathematics or engineering declines over the two years at CEGEP as a direct consequence of science instruction and this decline was reported amongst high, as well as low, achievers.

Seymour and Hewitt (1997) found that poor teaching was mentioned by more than 90% of switchers (p. 146), and that the atmosphere in science, mathematics and engineering classes was perceived as being cold compared with classes in the humanities. They further found that students equated good teaching with "openness, respect for students, encouragement of discussion, and the sense of discovering things together" (p.148). These qualities are consistent with the Fourteen Learner-Centered principles of student learning processes with which the American Psychological Association summarized research findings about good teaching and learning (American Psychological Association, 1997).

Despite this body of knowledge, Kardash and Wallace (2001) found that undergraduate science majors in biology and physics: a) see their classes as placing them in the role of passive learners; b) perceive grades, rather than an explanation of science concepts, to be the primary form of feedback provided; c) are unsure whether laboratory experiences support inquiry and problem-solving, as opposed to simply stating the correct answer; and d) find that faculty are seldom interested in them as students, and are instead preoccupied with their discipline. They further reported that 74% of the students who do persist in science, mathematics or engineering complain about poor teaching.

Seymour and Hewitt report that "traditional science pedagogy is inherently disadvantageous to women," (p. 235) and that females' preference for more cooperative learning experiences do not serve them well in the competitive ethos of science classes and contribute to their lower persistence compared with their male peers. More recently Epstein (2006) has pointed out that many students are forced to "slog through" two or more years of large, formulaic and impersonal introductory classes before they are introduced to the hands-on work that make a career in science attractive to working scientists.

To better understand how student and faculty perceptions of learning environments interact with student characteristics to impact on student academic performance and persistence in science, this study sought to understand students' perceptions of the learning climate in their classes. Further, teachers' conceptions of teaching were examined because these conceptions will determine how teachers translate their intentions into actions in their classrooms.

Prior research has demonstrated the relationship between students' conceptions of learning, their approach to learning tasks and learning outcomes (Kember & Kwan, 2002). Similarly, conceptions of teaching influence the approach to teaching adopted by faculty and in turn impact on the approach students adopt and the quality of the learning outcomes (Kember, 1997; Marton & Säljö, 1984). Entwistle and Walker (2002) explored the ways in which an ordered set of teachers' conceptions of teaching, from teacher-centered to student-centered, was associated with student learning outcomes, and concluded that the intellectual development of students was enabled by a student-centered approach to teaching rather than by a teacher-centered approach.

In 1997 Kember undertook a reconceptualization of research on the conceptions of teaching held by faculty in post-secondary education. All of the studies examined found distinct conceptions of teaching that could be arranged in an ordered set of qualitatively different conceptions from teacher-centered/content-oriented to student-centered/learning-oriented. After reviewing the studies he concluded that the outcomes could be synthesized into a two-tiered model, and that the teacher-centered/student-centered axis was a useful orientation for the ordered set (Figure 2).

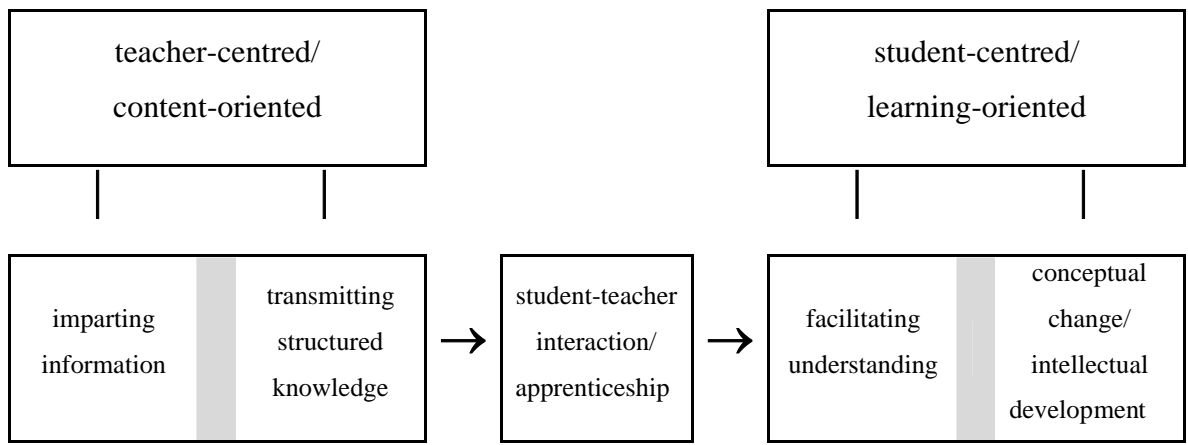

Figure 2: A two level model for categorizing conceptions of teaching.
The shaded areas indicate a diffuse boundary implying an easy development from one conception to the other. (after Kember, 1997)

Kember placed the conceptions under two broad orientations: The first is teacher-centered and focuses on the communication of specific content. The second is student-centered and focuses on students and learning. Each orientation is divided into two subordinate conceptions while a fifth transitional conception, where student-teacher interaction is first recognized as necessary, serves as a bridge between the two orientations. Rather than having well-defined boundaries the five conceptions of teaching are better thought of as well-established positions along a continuum (Prosser, Trigwell & Taylor 1994). After reviewing a number of frameworks (Kember, 1997; Kane, Sandretto & Heath, 2002; Havita & Goodyear, 2002) it was decided that that of Scardamalia and Bereiter (1989) was the most promising for the project. Their framework is non-judgmental, an important consideration when describing the project to faculty, it also describes classroom strategies and the mental gymnastics that faculty might use to overcome or avoid the deficits of each of the conceptions. Their four, "time-honored," (p. 37) conceptions are; *teaching as cultural transmission*; *teaching as skills training*; *teaching as the fostering of natural development*; and *teaching as producing conceptual change*. They consider the stages of their taxonomy to be hierarchical.

For each of these conceptions of learning the team developed a grid that described the actions of the teacher, descriptions of likely teaching activities, faculty expectations of student actions both inside and outside of the classroom and types of assessments used. As examples, for faculty who conceptualized teaching as transmission of knowledge or as fostering conceptual change, the characteristics are:

Teaching as transmission of knowledge: Human knowledge can accumulate and be transmitted from generation to generation. The teacher a) defines the curriculum by what they see as the standard knowledge of the subject; b) lectures and provides notes; c) maintains control; d) expects the students to be a passive recipient of knowledge and e) assessments focus on the correct answer.

Teaching as conceptual change: Learning is transformative rather than merely cumulative. The teacher a) identifies those areas where prior knowledge structures might contradict new knowledge structures; b) uses class time for short lectures, interactive activities and discussions that challenge students to resolve conflicts between prior and present conceptual knowledge; c) expects students to participate in discussions; d) prepares problem sets that challenge students' prior conceptions; and e) uses assessments that require that students demonstrate both correct problem solving algorithms and verbal and/or written responses relating to prior conceptions.

Methodology

*Participants*

Student participants (N = 1425) represent a cohort of academically able students who graduated from High School (grade 11) in June 2003 and entered one of four public Anglophone CEGEPs (Collège d'enseignement général et professionnel, two-year colleges preceding a three-year university program) in the fall. Participants volunteered to respond to at least one of three surveys administered during the four semesters of their college studies. The first survey was administered in class in the first weeks of students' first semester (fall 2003). The second was administered in science classes in the first weeks of students' second semester, while the third was administered in science classes in the students' fourth semester (spring 2005) after they had either applied for admission to university in a specific field, or were thinking about their choice for future studies.

Faculty participants (N = 84; 39% female, 61% male) were pre-university science program faculty at the four CEGEPs participating in this study and made up 42% of the population of mathematics and science faculty at these colleges. The participants were members of four departments: biology (6%), chemistry (25%), mathematics (37%), and physics (32%). Approximately 42% of faculty held a doctorate, 51% held a masters degree, and 7% did not reveal their education. Faculty who had more than 15 years of teaching experience made up 61% of the sample.

*Student Data*

Achievement and demographic data of students were obtained from academic records.

*The first survey*

Each participant responded to one of four randomly distributed 100-item versions of a 130-item survey. Consequently, the sample size for assessment of each of the variables varies and is less than the full 1425. All participants responded to items assessing *Interest in Science careers* (18 items), *Motivation to Select Science Courses in High School* (6 items), *Perceptions of learning environments in mathematics and science classes* (22 items) and *Teacher evaluation* (5 items). The latter two sets of items were a selection of high-factor-loading items from a scale developed by Kardash and Wallace (2001) that assessed students' perceptions of the classroom environment. The second part of the questionnaire assessed *Socio-economic Status* (parental education and family income), *Ethnic Background* and *Student Motivational Characteristics*, *Self-efficacy Beliefs* (6 items developed by Dedic, Rosenfield, Alalouf, & Klasa (2004)), and *Affect*. A nine-item Affect scale was used (Emmons, 1992). Participants were asked to rate each item, based on how they felt during the past week, using a Likert scale that ranged from 1 to 5, with 1 representing very slightly and 5 representing extremely. The items were: joyful, unhappy, worried/anxious, enjoyment/fun, depressed, pleased, happy, angry/hostile, and frustrated. These scales have excellent temporal reliability and internal consistency (Diener & Emmons, 1984).

*The second survey*

This survey assessed change in *Interest in Science Careers* (13 items) and *Perceptions of Learning Environments* in science and mathematics courses taken during the first semester in CEGEP (45 items).

*The third survey*

This survey was administered after students had made their choice of university program and asked them for their choices and the reasons for making those choices.

*Faculty data*

During the winter 2004 semester participants volunteered to answer an 80-item survey which included 30 items that assessed: a) faculty perceptions of the learning environment in their classes; b) their coping strategies; and, c) their course preparation practices. Thirty-seven items from the second student survey were rephrased to develop these thirty items. For example, the item "*Teachers attempted to find out what students already know about a topic before presenting new or more advanced information in their classes.*" from the student survey, became "*I attempt to find out what students already know about a topic before presenting the topic,*" in the teacher survey. Most but not all items were transformed. Some items had to be omitted from the faculty questionnaire because the faculty version would have clearly elicited skewed responses. Other items were potentially offensive to faculty, even though not to students. For example, although it was reasonable to ask students if they perceived that their instructor did not treat them with respect, asking faculty if they treat their students with respect was certain to be offensive.

Results

*Student and Faculty Perceptions of Learning Environments*

In the second survey students responded to 37 items describing learning environments in CEGEP. Exploratory factor analysis was carried out on the data from one CEGEP and a three-factor model determined. Confirmatory factor analysis was carried out on data from the remaining three CEGEPs and confirmed the three-factor model. A similar analysis was used for the first student survey that asked about learning environments in high school and revealed a three-factor model.

Exploratory factor analysis and confirmatory factor analysis of the faculty data revealed a two-factor model: *teaching as transmission* and *teaching as fostering student development*. The data did not support the four-factor model implied in the Scardamalia and Bereiter (1989) model but rather the three conceptions, teaching as skills training, teaching as the fostering of natural development, and teaching as producing conceptual change coalesced into the single conception, fostering student development.

There were clear similarities between the student models for high school and CEGEP. In both cases, the first student factor describes a supportive environment where teachers make students feel confident and competent. Teachers stimulate students to think along with them while they are explaining concepts, and encourage students' independent thinking. This factor was labeled *Fostering*.

The second student factor describes the classroom use of collaborative strategies, an environment where peer collaboration is encouraged or structured into lesson plans. Collaborative learning environments have been shown to be effective in enhancing both student motivation and learning (Lou, Abrami, Spence, Poulsen, et al., 1996; Springer, Stanne, & Donovan, 1999). This factor was labeled *Collaborative*.

Items that load on factor 3 in the student model, and on factor 2 in the faculty model, are very similar. They describe practices of faculty whose conception of teaching is *Transmission of Knowledge*. Unfortunately, all three surveys included too few such items, and some had to be removed from the analysis because they were highly skewed.

Ultimately four items loaded on these factors. Nonetheless these items are describing environments where the focus is on delivery of knowledge, not students' learning. This factor was labeled *Transmitting*. Table 1 provides a side-by-side comparison of the items that contributed to these models.

| | High School students | CEGEP students | CEGEP teachers | |
|---|---|---|---|---|
| **Factor 1** | Teachers explained their ideas in a way that made sense. | The teacher explained ideas in a way that made sense to me. | | **Factor 1** |
| | Teachers tried to ensure that their students felt confident and competent ... | The teacher tried to ensure that students felt confident and competent in the course. | | |
| | Teachers attempted to find out what students already know about a topic before presenting the topic | The teacher attempted to find out what students already knew about a topic before presenting the topic. | I attempt to find out what my students already know before showing them a new method of solving a problem | |
| | Teachers encouraged me to think for myself. | The teacher encouraged me to think for myself. | I encourage students to develop their own methods for solving typical problems. | |
| | Lectures stimulated me to think along with the teacher, and to understand new ideas. | Classes stimulated me to think along with the teacher, and to understand new ideas. | | |
| | Teachers emphasized the understanding of concepts more than the remembering of formulas. | The teacher emphasized the understanding of concepts more than the remembering of formulas. | I try to help students connect to the curriculum material by finding topics that will be interesting to them. | |
| | Teachers gave good examples and practical applications of mathematical and scientific concepts. | The teacher spent considerable time helping us to improve our skills and learn new ones. | In this course I focus on changing students' understanding of many concepts that they thought they already understood. | |
| | Teachers treated students with respect. | The teacher made me feel that making mistakes is a normal part of learning. | | |
| | Teachers related information presented in their classes to math or other science classes. | When I made a mistake in solving a problem I still got a good grade as long as my method was correct. | I often use examples from other science disciplines in my class | |
| | | The teacher often taught us several ways of solving the same problem. | I often show several ways of solving the same problem. | |
| **Factor 2** | Teachers encouraged students to work together. | The teacher asked students to work together as a regular part of classes. | | **Factor 2** |
| | When teachers asked groups of students to discuss a topic, the discussion usually improved my understanding. | The teacher encouraged students to discuss ideas amongst themselves as a way to improve understanding. | I encourage students to discuss ideas amongst themselves as a way to improve their understanding. | |
| | Teachers encouraged students to participate in classroom discussions. | The teacher allowed class time for debates about interesting issues that we brought up | | |
| | Teachers promoted the idea of "discovering things together" with students in their classes. | The teacher encouraged students to work with their peers to promote a sense of mutual support. | I encourage students to work with their peers to promote a sense of mutual support. | |
| | Teachers gave a short lecture and then groups of students worked on problems or discussed topics. | | | |
| **Factor 3** | I spent most of my time in class copying the teacher's notes. | I spent most of my time in class copying the teacher's notes. | Students should spend most of their time in class taking notes ... | |
| | Teachers lectured most of the time. | The teacher lectured most of the time. | | |
| | Teachers assumed that students knew more about math and science than they really do. | The teacher was mostly concerned with covering all the material that he planned to cover. | In my class I am mostly concerned with covering all the material that I planned to cover. | |
| | Group work in my classes mostly involved repetition of problems where one "plugs-in" numbers into a formula. | To succeed in this course I often had to memorize solutions. | I limit the number of questions I allow in a given class because of time needed to cover the material. | |

Table 1. Comparison of perceptions of the learning environments

Students' factor scores on *Fostering* and *Collaborative* correlated positively with students' perception that the teacher was effective in making them learn while students' factor scores on *Transmitting* correlated negatively with their perception of teachers' effectiveness as shown in Table 2.

Table 2. Correlations between students' perception of a class and student's perception that the teaching was effective in making them learn

|  | High School Students | College Students |
|---|---|---|
| Factor 1  Fostering | .589** | .819** |
| Factor 2  Collaborative | .307** | .364** |
| Factor 3  Transmitting | - .199** | - .359** |

Note  ** significance at level p = .01

*Persistence and Academic Performance*

A student's potential to succeed in science was measured by averaging their grades in high school science courses. Students whose score was 70% or greater had the ability to succeed in the science program (Dickie, 2000). There were N = 1425 students in this cohort (765 females, 660 males). There were no significant differences in the potential of females and males to succeed in science as measured by their high school science grades, (82.2% vs. 82.1%, respectively) nor were there differences in their performances in the first semester in CEGEP (83.6% vs. 82.2%). However at every stage of decision making about future careers in science, female students were more likely to abandon science (Figure 3).

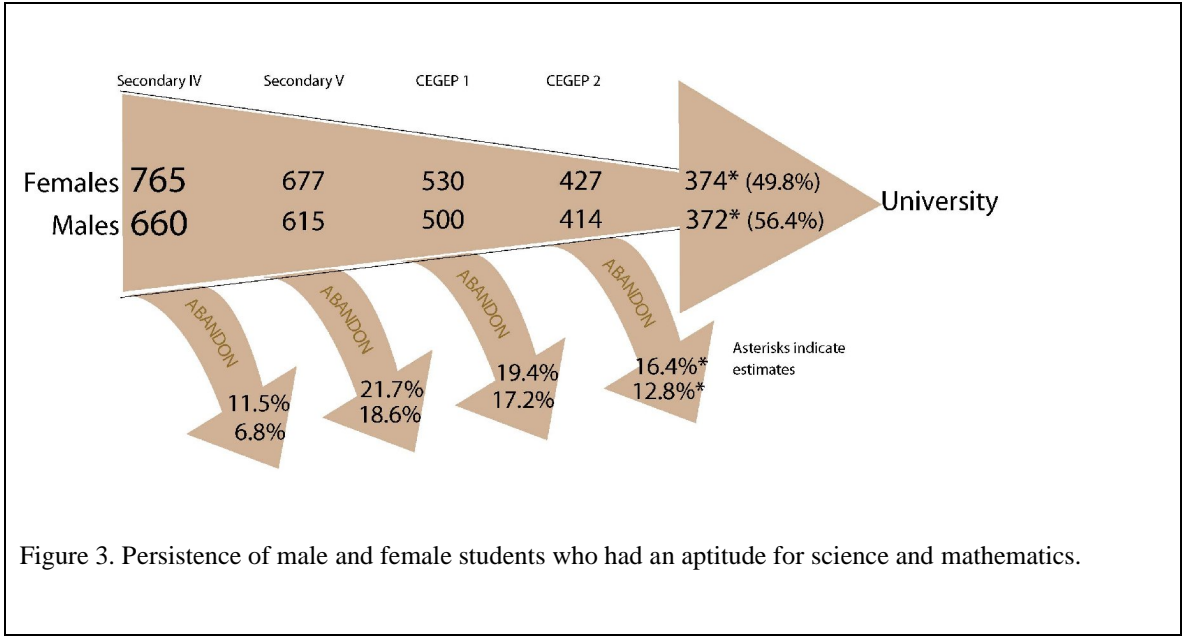

Figure 3. Persistence of male and female students who had an aptitude for science and mathematics.

There was a significant difference between the science potential of those who persisted in science at CEGEP and those who abandoned science after high school but this difference is small (83.6% vs. 82.2%, persist/abandon) and is unlikely to have practical impact. For those who persisted there was a small but non-significant difference in the average CEGEP grade over three semesters in favor of females (73.1% vs. 72.2%, N = 841, p = .059). Female students in this cohort of academically able students do perform better but can such a small difference in average grades have any practical meaning?

*Persistence and Attitude, Self-efficacy and Perception of the Learning Climate*

Grades alone do not provide explanations for why capable students abandon science, or why bright female students are more likely to forgo science careers than their male peers. Do their attitudes and beliefs provide clues? It was determined that there were statistically significant differences in females' and males' attitudes towards science, means 3.7 vs. 3.9, (F(1,1425) = 13.4, p < .001) on a scale from 1, "I hate sciences and mathematics" to 5, "I love sciences and mathematics". This difference points to students' beliefs about science as a possible contributor to their abandoning science.

To further explore this difference in feelings toward science, students' positive feelings and the strength of their self-efficacy beliefs were assessed first as they entered CEGEP and again, after completion of the first semester. In the cohort of 1425 students, data were available for 398 students. There was no significant difference between the feelings towards science of females and males at the beginning of CEGEP, when they were reporting on their experiences in high school, (means 3.6 vs. 3.6 respectively, $F_{(1,398)}$ = .8, p = .36). However, after one term at CEGEP, there was a significant difference, (means 2.9 vs. 3.2, $F_{(1,398)}$ = 15.1, p < .001) with males feeling more positive towards science, as shown in Figure 4.

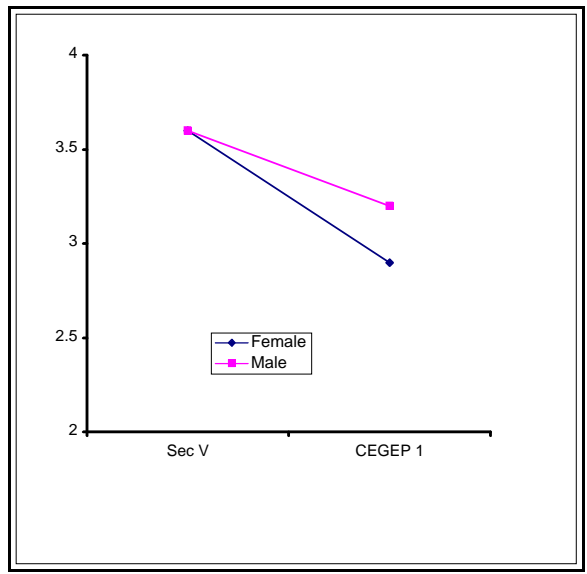

Figure 4.  Means of the frequency of positive feelings for female and male students. (scale: 2 = "just a few times"; 3 = "often"; 4 = "quite often")

Similarly, there were significant differences in self-efficacy beliefs between females and males, both at the beginning of CEGEP (when they were reporting on their experiences in high school) and after the first semester of CEGEP (HS: means 3.4, 3.7 for females and males respectively, $F_{(1,398)}$ = 42.6, p < .001; CEGEP: means 3.2, 3.5 for females and males respectively, $F_{(1,398)}$ = 19.9, p < .001). Both females and males become less confident in their abilities to tackle tasks after one term of studies in the science program. Males are more confident than females and the difference between female and male self-efficacy beliefs, initially seen in high school, is repeated, albeit at lower values, after the first term of CEGEP (Figure 5). The differences and changes in self-efficacy beliefs during the transition period are not sufficient to explain why female students are leaving science in greater numbers than their male peers.

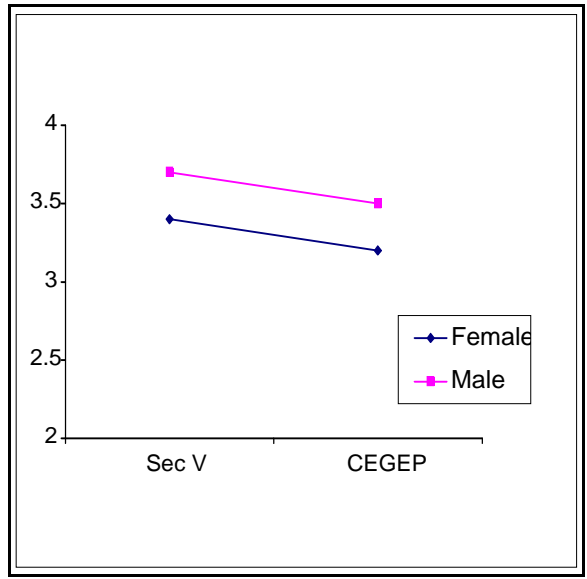

Figure 5. Self-efficacy beliefs after Secondary V and after the first semester of CEGEP

The greater self-efficacy of males entering CEGEP compared to females might be an example of males systematically overestimating their own ability and of females underestimating their ability (Sax, 2005, p. 43). Pajares (2003) has suggested that males are more likely to express confidence in skills they may not possess and to express overconfidence in skills they do possess.

Students' perceptions of learning environments were also examined. There were no significant gender differences in the perception of the learning environments that are student-friendly and nurture their learning (Fostering). However, it was found that female students perceive the learning environment as more teacher-centered (Transmitting) than their male peers both in high school, (mean 3.0, 2.8 females/males respectively, F(1.398) = 7.1, p = .008), and in CEGEP, (mean 3.4, 3.2, F(1.398) = 7.7, p = .006). Both genders perceived the learning environments in CEGEP as more Transmitting than in high school (Figure 6).

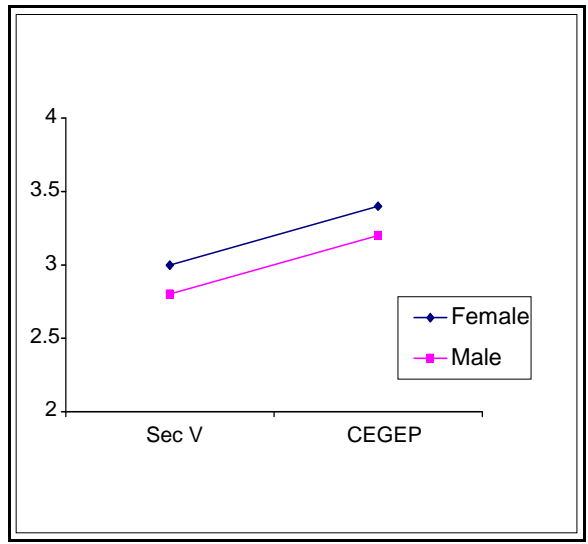

Figure 6. Perceptions of teacher-centred transmitting learning environment

It is important to note that the perception of a supportive environment (Fostering) positively correlated with students' perception of teacher effectiveness. In addition, there was also a significant difference in academic performance between students who saw their classes as more-effective/less-effective in promoting learning (means 83.0, 81.3), $F(1, 1302) = 16.7$, $p < .001$).

## Discussion

The most important finding of this study is that for the cohort of academically able students, ability (as measured by high school performance) and performance (as measured by CEGEP grades) cannot explain the difference in persistence of females and males. This inability of performance to predict persistence is in agreement with the findings of Fehrs and Czujko, (1992) who found that females who left physics performed on par with males who persisted, but is at odds with the findings of two large studies of gender differences in science, mathematics and engineering graduates that were reviewed by Seymour and Hewitt (1997); these two studies reported that similar performance lead to similar persistence (Strenta, Elliott, Adair, Matier et al., 1994; Ginorio, Brown, Henderson & Cook, 1993).

The persistence and success of students cannot be divorced from the context in which they are embedded. The four colleges are commuter colleges and the science program allows students only a limited choice of optional courses. Tinto (1997) has pointed out that in commuter colleges, students' academic interactions with other students and with faculty take place in the classroom. Indeed the classroom might be the only context where this academic engagement takes place so it is crucial to success and persistence. In the science program there is a compulsory core of courses in Biology, Chemistry, Physics and Mathematics. Parker, Rennie, and Harding (1995) in a meta-analysis, found that in every country for which data were available the participation levels of females in science beyond the compulsory courses was lower than that of males. This effect was particularly notable in Physics, less so in Chemistry and small for Biology. They also point out that when they are free to choose, "females and males participate quite differently, (in both quantitative and qualitative terms) in the study of science" (p. 188).

Seymour and Hewitt suggest that a clue to finding the cause of the poor persistence of females is that women in both graduate and undergraduate levels reported that feelings of psychological alienation were important factors in their abandoning science. In this study we found differences in the perception of the learning climate. In particular, females perceived classes as more teacher-centered (Transmitting) than did their male counterparts.

We suggest that this difference in the perception of the learning climate, combined with their lower feelings of self-efficacy are important contributors to the poor persistence of females. In making this hypothesis we are in agreement with the findings of Kubanek and Waller (1996) who followed a randomly selected sample of women students for four semesters from when they entered CEGEP until they graduated, abandoned science or continued their science studies. They too found that their quantitative data, from the Coopersmith Self-Esteem Inventory (1984), showed no relationship between either self-esteem and persistence, or self-esteem and performance (as measured by CEGEP grades). They did however find that a student's perception of how their questions were received (both in-class and out-of-class) was crucial to their self-esteem and confidence. Female students' perception that their questions were not encouraged, and their perceptions of how they could or could not relate to their teachers, were associated with abandoning science.

This study found a connection between self-efficacy, students' perception of the ability of the teacher to help them learn, and persistence. Female students who were in the "more effective in making be learn" group have higher self-efficacy than their female counterparts in the "less effective in making me learn" group and are more likely to persist. Male students in either group have higher self-efficacy than females. Perhaps the male students have developed an "immunity" to effects of the learning environment on self-efficacy, and hence persistence. This suggestion is supported by the report of Tobias (1990) who reviewed a study of 4,000 Ph.D. scientists and engineers in NASA (Dietz, Lund & Rosendhal, 1989), where it was found that "over 80% decided on a career in science or engineering before completing high school," and that "the intrinsic interest of the subject matter" was more important than, "all other influences," including their high school and college teachers, in their decision to study science or engineering (Tobias, 1990, p. 10).

One key factor in classes that are perceived as Fostering is the involvement of students beyond silent listening and copying notes. When students perceived classes as supportive of their engagement, their academic performance was better. On the other hand, if the climate of a class is perceived as Transmitting, persistence is negatively affected: more so for women. While "Transmitting" and "Fostering" are convenient constructs for characterizing the learning climate created by the teacher the expectations of women and men can be teased out to more fully describe the desired faculty-student relationship. Men have described a "good" teacher as one who is "enthusiastic," "interesting," "entertaining," and "can explain well," while women characterize a "good" teacher as one who is "approachable," "friendly," "patient," and "interested in how you respond" (Seymour and Hewitt, 1997, p. 305). While a Harvard study (Light, 1990, cited in Powell, 2005) found that men wanted an advisor who "knows the facts," or one who "makes concrete and direct suggestions, which I'm then free to accept or reject." In comparison women wanted an advisor who "will take the time to get to know me personally," or who "is a good listener and can read between the lines if I am hesitant to express a concern." Kubanek and Waller (1996) avoided comparing the perceptions of men and women but did find that women students sought to establish a relationship with their teachers and used the teacher's response to their questions as a barometer of whether a relationship had been established.

This study found that teachers who are perceived as more Fostering and less Transmitting by their students (and so enable better performance and persistence), are more likely to have a degree or diploma in Education, in agreement with the findings of Darling-Hammond et al. (2005) who found that teacher preparation did matter and resulted in higher performance of elementary students. The lack of training in pedagogy of most CEGEP faculty is unfortunate when, in response to an open-ended question asking for obstacles to their teaching, 52% of teachers' comments concerned students' lack of academic preparation in study skills and strategies, poor written and oral communication skills, lack of deep approaches to learning, and poor student motivation and interest. These student 'knowledge of how to participate in higher education' factors and motivation issues have a rich empirically based literature that point to avenues of intervention in mathematics and sciences by the regular classroom teacher (Arons, 1990; Ramsden, 1992; Redish, 2003; Sultan & Artz, 2003).

*Conclusions and Recommendations*

Science faculty (at high school, college and university) must be made aware of the relationship between students' perceptions of the learning climate in their classes and persistence and success, and of the different perceptions of their actions by male and female students. Unless the teacher sets out to create a Fostering environment it is unlikely that students in that class will perceive it as such. In practical terms this means less lecturing and more interactive-engagement. Faculty must also be challenged as to why they are rigorous in their research but rely on intuition when they teach. A consequence of this lack of rigor is that, "unlike most physics problems, problems in education do not stay solved" (Hehn & Neuschatz 2006 p. 38).

"Fostering" classes involve more student-student interactions and less lecturing and quiet copying of notes. For several years this teacher-talking, student-listening model of instruction has been known to be less effective than active learning in improving students' conceptual understanding of physics and other disciplines (Hake, 1998). There have been recent and notable articles in high status journals like *Science* and *Nature* pointing out the failures of much of traditional teaching, and the promise of active learning (Handelsman et al., 2004; Powell, 2003). Teaching with interactive strategies not only yields significantly increased understanding for both males and females, but also reduces the gender gap in conceptual understanding in physics (Lorenzo, Crouch, & Mazur, 2006) and benefits historically underserved minority and marginalized students (Kuh, Kinzie, Cruce, Shoup, & Gonyea, 2006).

The multi-year study of Tai, Liu, Maltese, & Fan, (2006) found that students who expressed an interest in a science career early in their schooling were more likely to persist and earn science, engineering or technology degrees. This suggests that encouraging students in science at an early age can play a significant role in their career choices. Unfortunately there is little science done in elementary schools in Quebec (Lenoir, Larose, & Geoffroy, 2000), so change must begin in elementary schools if a high level of interest in science on the part of students, both girls and boys, but especially girls, is to be maintained into high school, college and beyond. Intervention must begin by ensuring that teachers at elementary and high schools have experience with the methodology of science so that they can feel comfortable using it in their classrooms. Intervention need not be confined to the classroom however, as is shown by after-school programs that use, for example, community gardens and recycling projects to develop understanding of the methodology of science. *Les Scientifines*, an after-school program serving elementary school girls from a low-income community in Montreal, offers not only a safe place for a snack and homework but activities, for example, a project on rockets for a science fair, that develop girls' scientific literacy and understanding of the scientific method (Rahm, Moore, & Martel-Reny, 2005). Further, such programs expose children from low-income, frequently immigrant, communities "to a variety of financially interesting career paths potentially new to them, that if pursued, could help them break out of the vicious cycle of poverty" (Chamberland, Théoret, Garon, & Roy, 1995). Currently the majority of CEGEP science faculty are discipline experts, with little or no knowledge of education research. If student persistence and success in science in CEGEP are to be increased then we must take the time and energy to educate faculty about research on teaching and learning, and in particular about ways in which their teaching can become more effective. We must not allow them to be seduced by the tendency to blame problems on other milieu (e.g., high school), and accept instead that changes in their own practices are required and can enable real empowerment of learners. This empowerment can extend beyond the (justifiably) immediate concern of faculty with increasing students' understanding of their own particular discipline, to appreciating the important contribution they as faculty can make to advancing wider societal goals like reducing inequality, unemployment and poverty, and increasing economic well-being by advancing persistence, success and science literacy.